\documentclass[aps,preprintnumbers,eqsecnum,amsmath,amssymb,nofootinbib]{revtex4}
\usepackage{graphicx}
\usepackage[english]{babel}
\usepackage{url}
\usepackage{graphicx}
\usepackage{xcolor}
\usepackage{amsmath}
\usepackage{amssymb}
\usepackage{slashed}
\usepackage{multirow}
\begin{document}
\title{On the Resonance Coupling and Width in Quantum Field Theory}
\author{Dmitri Melikhov$^{a,b,c}$}
\affiliation{
$^a$D.~V.~Skobeltsyn Institute of Nuclear Physics, M.~V.~Lomonosov Moscow State University, 119991, Moscow, Russia\\
$^b$Joint Institute for Nuclear Research, 141980 Dubna, Russia\\
$^c$Faculty of Physics, University of Vienna, Boltzmanngasse 5, A-1090 Vienna, Austria}
\date{\today}
\begin{abstract}
  In quantum field theory, characteristics of resonances are related to self-energy diagrams,
  which are ultra-violet divergent and require renormalization. We demonstrate the proper way to
  define the resonance coupling $g_M$ such that the resonance properties calculated in quantum
  field theory are finite and scheme-independent quantities. 
\end{abstract}
\maketitle

\section{Introduction}
Resonance characteristics such as mass, width, and couplings are usually extracted from the
experimental results for physical observables using some theoretical formulas.
On the theoretical side, quantum field theory (QFT) provides the basis
for the description of the resonance properties. Feynman diagrams of QFT contain ultraviolet (UV)
divergences and require renormalization such that the renormalized couplings and masses in the Lagrangian are scheme-dependent quantities -- scheme-dependence emerges due to different possibilities of fixing finite subtraction polynomials
in the UV-divergent diagrams. As a result, relating scheme-dependent quantities of QFT to the
physical resonance parameters requires some care.

In this letter, we revisit resonance contributions to scattering amplitudes in QFT and show the proper way of relating the renormalized scheme-dependent resonance parameters of QFT and the
physical resonance parameters. 

An essential ingredient of our discussion is the UV divergence of the
self-energy Feynman diagram. We therefore consider examples of the effective Lagrangians leading 
to quadratically and logarithmically divergent self-energy diagrams. 

\section{Resonance contribution to the $\bar\chi\chi$ scattering amplitude}

Let us start with a simple model of the interaction of two fermions $\bar\chi\chi$ with a scalar
resonance $R$ described by the effective Lagrangian 
\begin{eqnarray}
  \label{Leff1}
L_{\rm eff}=g\,\bar\chi\chi R. 
\end{eqnarray}
Our discussion in this Section in its essential features will apply to a generic triple interaction of the type (\ref{Leff1}).

\begin{center}
\begin{figure}[h!]
  \includegraphics[width=11cm]{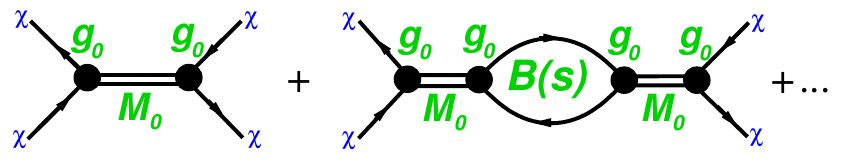}
  \caption{\label{Fig:1}The diagrams describing the series for $A$ in Eq.~(\ref{a1}).}
\end{figure}
\end{center}

The $s$-channel amplitude of the $\bar\chi\chi$ interaction via a scalar resonance $R$ after the summation of the self-energy diagrams, Fig.~\ref{Fig:1}, takes the following form \cite{mn}:
\begin{eqnarray}
  \label{a1}
  A(s)=\frac{g_0^2}{M_0^2-s-g_0^2 B(s)}, \qquad B(s)={\rm Re}\,B(s)+i\,{\rm Im}\,B(s). 
\end{eqnarray}
Here $g_0^2 B(s)$ is an UV divergent one-loop self-energy diagram with $\bar \chi\chi$ fermion pairs
in the loop and a bare scalar vertex $g_0R\bar\chi\chi$. The coupling $g_0$ and the resonance mass
$M_0$ are infinite bare parameters. 

The imaginary part ${\rm Im}\,B(s)$ can be calculated without ambiguity, but the dispersion representation for
its real part ${\rm Re}\,B(s)$ needs two subtractions. 
After renormalization, ${\rm Re}\,B(s)$
contains an arbitrary subtraction polynomial of the first order in $s$ and thus two arbitrary constants. 
These constants may be fixed in different ways \cite{mn,anisovich,hagop,gs}; the corresponding ambiguity is called the scheme dependence. 

So, after renormalization, the amplitude takes the form 
\begin{eqnarray}
  \label{a2}
  A(s)=\frac{g_R^2}{M_R^2-s-g_R^2\,B_R(s)}, \qquad B_R(s)={\rm Re}\,B_R(s)+i\,{\rm Im}\,B_R(s),  
\end{eqnarray}
where $g_R$ is finite but scheme-dependent renormalized coupling constant,
${\rm Re}\,B_R(s)$ is finite but contains arbitrary finite parameters to be fixed by the choice of the scheme, whereas
${\rm Im}\,B_R(s)={\rm Im}\,B(s)$ is finite and contains no scheme-dependence (i.e., no ambiguity).
Hereafter we use for the imaginary part the notation ${\rm Im}\,B(s)$. 

Let us introduce the physical parameters: the resonance mass $M$, the resonance width $\Gamma$, and the 
fermion mass $m$. [$M_R$ is related to the physical mass 
$M$ but is not identical to it; see Eq.~(\ref{M})]. 
If we consider the case $M<2m$, then the resonance is stable and corresponds to the pole on the real axis below the two-particle threshold at $s=4m^2$. If the mass parameter satisfies the relation 
$M > 2m$ the resonance acquires width $\Gamma$. One may try to define the width as  
\begin{eqnarray}
\label{g}
\Gamma=g^2_R \frac{{\rm Im}\, B(M^2)}{M}.  
\end{eqnarray}
However, this would be an inconsistent idea, which leads to a problem:
The imaginary part of the self-energy diagram in one loop is scheme-independent,
but $g_R$ in Eq.~(\ref{a2}) depends on the scheme. Therefore, $\Gamma$ defined by (\ref{g}) is scheme-dependent and cannot
be a physical characteristic of a resonance. Obviously, a scheme-dependent coupling $g_R$ cannot appear in the physical width and should be replaced by some scheme-independent coupling. We shall see that a proper definition of the resonance
coupling requires a proper treatment of the real part of the self-energy diagram.

We neglect for the moment ${\rm Im}\, B(s)$ (we shall restore it later) and keep the real part ${\rm Re}\,B(s)$. Then the amplitude $A$ has a pole at $s=M^2$ and can be written as 
\begin{eqnarray}
\label{a3}
A=\frac{g^2_M}{M^2-s}+\dots,   
\end{eqnarray}
where $\dots$ denote regular terms for $s\to M^2$. Here, $g_{M}$ is a properly defined resonance coupling. 
The resonance width then takes the form 
\begin{eqnarray}
\label{g1}
\Gamma =g^2_M\frac{{\rm Im}\,B(M^2)}{M}.   
\end{eqnarray}
We shall see that the appearance of the coupling $g_M$ is accompanied by a
specific subtraction in ${\rm Re}\,B(s)$. Moreover, we shall see that $g_M$ defined in Eq.~(\ref{a3}) as 
the residue in the pole at $s=M^2$ does not depend on the renormalization scheme and therefore represents
the physical resonance coupling.

\section{\label{I}
Quadratically-divergent Self-energy and the resonance coupling}
For the case of a scalar-resonance interaction with two fermions, described by the effective Lagrangian (\ref{Leff1}),
the self-energy diagram $B(s)$ diverges quadratically in the UV \cite{blm2025a}. 
Using the language of dispersion representations, this means that
${\rm Im}\,B(s)$ is unambiguous, but the dispersion representation for ${\rm Re}\,B(s)$ needs
two subtractions and contains
an arbitrary polynomial of first order in $s$. Let us fix this subtraction polynomial in some way (i.e., choose a specific scheme), for instance, let us set $B(s=0)=0$ and $B'(s=0)=0$ and find \cite{mn}: 
\begin{eqnarray}
\label{bR}
B_R(s)&=&s^2
\;\int\limits_{4m^2}^\infty \frac{ds'}{\pi(s'-s-i0)s'}\left(1-\frac{4m^2}{s'}\right)^{3/2}. 
\end{eqnarray}
In this scheme, we denote the resonance mass parameter as $M_R$ and the coupling as $g_R$ and the amplitude takes the form
(\ref{a2}): 
\begin{eqnarray}
  \label{2.a1}
  A=\frac{g_R^2}{M_R^2-s-g_R^2 B_R(s)}. 
\end{eqnarray}
Let us introduce the physical resonance mass $M$ by the relation 
\begin{eqnarray}
\label{M}
M_R^2=M^2-g_R^2 B_R(M^2). 
\end{eqnarray}
Then, in the stable resonance case $M<2m$, the amplitude (\ref{2.a1}) has a pole at $s=M^2$ but
the expansion of the amplitude near this pole does not have the required form (\ref{a3}). 
To reduce the resonance contribution to the form (\ref{a3}),
let us introduce the resonance coupling $g_M$ as follows 
\begin{eqnarray}
 g_M^2=\frac{g_R^2}{1+g_R^2\,{\rm Re}\,B'(M^2)}.  
\end{eqnarray}
Using this coupling, we can rewrite $A$ as follows:
\begin{eqnarray}
  \label{2.a2}
  A=\frac{g^2_M}{M^2-s- g^2_M{\rm Re}\,B_R(s,M^2)-i g^2_M{\rm Im}\,B(s)},
  \end{eqnarray}
where   
\begin{eqnarray}
{\rm Re}\,B_R(s,M^2)\equiv {\rm Re}\,B_R(s)-{\rm Re}\,B_R(M^2)-{\rm Re}\,B'_R(M^2)(s-M^2).  
\end{eqnarray}
The quantity ${\rm Re}\,B_R(s,M^2)$ behaves like $O\left((s-M^2)^2\right)$, so the expansion of (\ref{2.a2}) near $s=M^2$ 
has the desired form (\ref{a3}). This means that now the resonance width takes the form 
\begin{eqnarray}
  \label{2.g}
  \Gamma=g_M^2\frac{{\rm Im}\,~B(M^2)}{M}. 
\end{eqnarray}
Here, ${\rm Im}\,B(M^2)$ is scheme-independent, while the scheme-independence of $g_M$ may not be immediately obvious.
So let us turn to the expression for the amplitude (\ref{2.a2}):
The subtracted real part ${\rm Re}\,B_R(s,M^2)$ corresponds to the renormalized self-energy diagram from which two first terms of its Taylor expansion at the point $s=M^2$ are subtracted.
Recall that the scheme-dependence of $B_R(s)$ reduces to the choice of the coefficients in 
the ambiguous polynomial of first order in $s$ contained in $B_R(s)$. It is easy to check that this polynomial completely falls out of ${\rm Re}\,B_R(s,M^2)$ which is thus a scheme-independent quantity.

In the end, the expression (\ref{2.a2}) contains no scheme-dependence, and so $g_M$ is a {\it finite scheme-independent
physical resonance coupling}.

\section{\label{II}
Logarithmically-divergent Self-energy and the resonance coupling}
From the previous Section we could get an impression that the difference between the renormalized coupling $g_R$ and the resonance coupling $g_M$ has its origin in the quadratic divergence of the self-energy diagram. We shall now see that this is not the case. 

Let us consider an example of the theory where the coupling is not renormalized at all. This is an interaction of a scalar particle $\phi$ with a scalar resonance $R$ described by an effective Lagrangian discussed in \cite{hagop}:
\begin{eqnarray}
L_{\rm eff}=g' \phi^+\phi\, R.
\end{eqnarray}
The one-loop self-energy diagram in this case is logarithmically divergent; it may be written as a dispersion representation with one subtraction  
\begin{eqnarray}
\label{bRs}
B_{Rs}(s)=s
\;\int\limits_{4m^2}^\infty \frac{ds'}{\pi(s'-s-i0)s'}\left(1-\frac{4m^2}{s'}\right)^{1/2} 
\end{eqnarray}
and contains one arbitrary subtraction constant. As noted in \cite{hagop}, this subtraction constant
may be absorbed into the renormalized resonance mass $M_R$ so that the coupling $g'$ does not
undergo any renormalization. The amplitude then takes the form
\begin{eqnarray}
\label{4.3}
A(s)=\frac{g'^2}{M_R^2-s-g'^2 B_{Rs}(s)}. 
\end{eqnarray}
Intuitively, one may expect that, since the coupling $g'$ is finite and does not require renormalization, this very coupling is the proper resonance coupling. However, this is not the case!

The key point is that the properly defined resonance coupling is related to the residue in the pole at $s=M^2$ by (\ref{a3}). In order to obtain the behavior of the amplitude near the pole of the form (\ref{a3}), we have to perform the redefinition of the coupling precisely in the same way as in the case of the quadratically divergent $B_R(s)$ of Sec. \ref{I}.
That is, in the denominator of (\ref{4.3})  we have to isolate 
the combination
\begin{eqnarray}
B_{Rs}(s,M^2)=B_{Rs}(s)-B_{Rs}(M^2)-B_{Rs}'(M^2)(s-M^2), 
\end{eqnarray}
which behaves as $O\left((s-M^2)^2\right)$ near the pole, and rewrite the amplitude in the form  
\begin{eqnarray}
  \label{b2}
  A=\frac{g'^2_M}{M^2-s- g'^2_M{\rm Re}\,B_{Rs}(s,M^2)-i g'^2_M{\rm Im}\,B(s)},
  \end{eqnarray}
with 
\begin{eqnarray}
 g'^2_M=\frac{g'^2}{1+g'^2 {\rm Re}\,B_{Rs}'(M^2)}.  
\end{eqnarray}
The resonance coupling $g'_M$ differs from $g'$ despite the fact that the coupling $g'$ is finite and does not undergo renormalization. This example clearly shows that, in general, the physical resonance coupling $g_M$ differs from 
any (even finite) coupling in the Lagrangian. 

\section{Conclusions}
In this letter, we revisit a proper way to define the resonance coupling and its width in QFT. 

\vspace{.5cm}
\noindent
(i) The properly defined resonance coupling $g_M$ is related to the residue 
in the pole (for $M<2m$ when the resonance is stable) of the scattering amplitude as follows
\begin{eqnarray}
    A(s)=\frac{g_M^2}{M^2-s}+\dots 
\end{eqnarray}
where $\dots$ denote regular terms for $s\to M^2$.

\vspace{.5cm}
\noindent
(ii) The scattering amplitude has the following parametrization in terms of the physical resonance coupling $g_M$ and the resonance self-energy diagram: 
\begin{eqnarray}
  \label{c1}
A(s)=\frac{g^2_M}{M^2-s-g^2_M\, {\rm Re}\,B_R(s,M^2)-i g^2_M{\rm Im}\,B(s)]},  
\end{eqnarray}
where
\begin{eqnarray}
\label{c2}
  {\rm Re}\,B_R(s,M^2)={\rm Re}\,B_R(s)-{\rm Re}\,B_R(M^2)-{\rm Re}\,B'_R(M^2)(s-M^2).  
\end{eqnarray}
Any renormalization scheme for a {\it scheme-dependent} $B_R(s)$ may be used to build ${\rm Re}\,B_R(s,M^2)$ in (\ref{c2}).  
The quantity ${\rm Re}\,B_R(s,M^2)$ as well as the coupling $g_M$ are finite and {\it scheme-independent}.
The appearance of a scheme-independent coupling $g_M$ accompanies the
appearance of a scheme-independent ${\rm Re}\,B_R(s,M^2)$ in (\ref{c1}). 

It should be emphasized that the appearance of a subtracted $B_R(s,M^2)$ in the amplitude (\ref{c1}) {\it is not related} to a divergent behavior of the self energy diagram in the UV region. 
The order of the $s$-polynomial subtracted from $B_R(s)$ in (\ref{c2}) is not related to the order of the UV divergence of the self-energy diagram: 
One and the same $B_R(s,M^2)$ emerges in the case of a quadratically-divergent self-energy of Sec.~\ref{I} and in the case of a logarithmically-divergent self-energy of Sec.\ref{II}. Even in the (purely hypothetical) case of a theory without any UV divergences and a finite self-energy, the physical resonance coupling would be different from the resonance coupling in the Lagrangian. 

\vspace{.4cm}
{\it Independently of the UV behavior of the self-energy, the physical resonance coupling $g_M$ is different from the resonance coupling in the Lagrangian. The proper parameterization of the amplitude in terms of $g_M$ has the form (\ref{c1}) with the subtracted ${\rm Re}\,B_R(s,M^2)$ of (\ref{c2}).} 

\vspace{.5cm}
\noindent
(iii) The physical resonance width is given by 
\begin{eqnarray}
  \label{c3}
  \Gamma=g^2_M\frac{{\rm Im}\,B(M^2)}{M}. 
\end{eqnarray}
This is a finite {\it scheme-independent} quantity. 

\vspace{.5cm}
The expressions (\ref{c1})-(\ref{c3}) should be used to extract parameters of broad resonances from the data. 

\acknowledgments
The author has the pleasure to thank M.~A.~Ivanov, O.~Nachtmann, and H.~Sazdjian for the illuminating discussions. 
The work was carried out within the program {\it Particle Physics and Cosmology} of the National Center of Physics and Mathematics. 



\begin{thebibliography}{50}%
\bibitem{mn}
D. Melikhov, O. Nachtmann, V. Nikonov, and T. Paulus,
{\it Masses and couplings of vector mesons from the pion electromagnetic, weak,
  and $\pi\gamma$ transition form-factors}, Eur.~Phys.~J.~C~{\bf 34}, 345 (2004).
\bibitem{anisovich}
A.~V.~Anisovich, V.~V.~Anisovich, and A.~V.~Sarantsev,
{\it Scalar Glueball: Analysis of the $IJ^{PC} = 00^{++}$-wave},
Z.~Phys.~A~{\bf 359}, 173 (1997).   
\bibitem{hagop}%
H.~Sazdjian, {\it The Interplay between Compact and Molecular Structures in Tetraquarks}, 
Symmetry {\bf 14}, 515 (2022).
\bibitem{gs}
G. J. Gounaris and J. J. Sakurai, 
{\it Finite width corrections to the vector meson dominance prediction for $\rho\to e^+e^-$}, 
Phys.~Rev.~Lett.~{\bf 21}, 244 (1968).

\bibitem{blm2025a}%
 A. Berezhnoy, W. Lucha, and D. Melikhov, 
  {\it Analysis of $q^2$-distribution for $B\to K M_X$ and $B\to K^* M_X$
    in a scalar-mediator dark matter scenario}, Phys.~Rev.~D {\bf 111}, 075035 (2025). 
\end{thebibliography}
\end{document}